\documentclass{article}

\usepackage{arxiv}

\usepackage[utf8]{inputenc} 
\usepackage[T1]{fontenc}    
\usepackage{hyperref}       
\usepackage{url}            
\usepackage{booktabs}       
\usepackage{amsfonts}       
\usepackage{nicefrac}       
\usepackage{microtype}      

\usepackage{graphicx}
\usepackage{amsmath}
\providecommand{\shortcite}[1]{\cite{#1}}

\title{Smart, Deep Copy-Paste}

\author{
  Tiziano Portenier \\
  University of Bern\\
  Switzerland \\
  \texttt{portenier@inf.unibe.ch} \\
   \And
 Qiyang Hu \\
  University of Bern\\
  Switzerland \\
  \texttt{hu@inf.unibe.ch} \\
   \And
 Paolo Favaro \\
  University of Bern\\
  Switzerland \\
  \texttt{favaro@inf.unibe.ch} \\
  \And
 Matthias Zwicker \\
  University Maryland, College Park\\
  United States of America \\
  \texttt{zwicker@cs.umd.edu} \\
}

\begin{document}
\maketitle

\begin{abstract}
In this work, we propose a novel system for smart copy-paste, enabling the synthesis of high-quality results given a masked source image content and a target image context as input. Our system naturally resolves both shading and geometric inconsistencies between source and target image, resulting in a merged result image that features the content from the pasted source image, seamlessly pasted into the target context. Our framework is based on a novel training image transformation procedure that allows to train a deep convolutional neural network end-to-end to automatically learn a representation that is suitable for copy-pasting. Our training procedure works with any image dataset without additional information such as labels, and we demonstrate the effectiveness of our system on two popular datasets, high-resolution face images and the more complex Cityscapes dataset. Our technique outperforms the current state of the art on face images, and we show promising results on the Cityscapes dataset, demonstrating that our system generalizes to much higher resolution than the training data.
\end{abstract}


\section{Introduction}
Image manipulation techniques consitute a major part in computer graphics and computer vision research. The rapidly growing imagery content on the web, accelerated by social media trends and high-quality image acquiring systems of modern smartphones increase the demand for flexible, high-quality, and easy to use image editing applications. Recently, several tools have been proposed that leverage deep learning techniques to manipulate and enhance images by inexperienced users, for example \cite{Yan2016, IizukaSIGGRAPH2016, gharbi2017deep, zhang2017real}. However, most applications target rather specific image editing operations and lack more general purpose image manipulation. On the other hand, image editing applications targeted for experienced users allow to enhance images by directly manipulating individual pixel values. While these systems allow complex editing operations, expertise is required to use such tools efficiently, and reasonably complex manipulations are time consuming even for experienced users.

In this work, we propose an image editing system that enables an user to manipulate images by using imagery content from another source image. We make use of recent advances in deep learning to implement a smart copy-paste system that produces realistic, high-quality results by seamlessly merging a source image patch into a target image. This approach allows to perform complex image manipulations given that there is another exemplar image that contains the desired features. In Section~\ref{sec:results} we show several examples where an object is copied from a source image and seamlessly pasted into a target image.

Conceptually, our work can be seen as a form of image completion and it is technically related to approaches that leverage deep learning to solve the image completion problem \cite{pathakCVPR16context, IizukaSIGGRAPH2017, Li_2017_CVPR, Yeh_2017_CVPR}. While these approaches try to inpaint a missing image region solely based on the image context to synthesize a plausible image patch, in our work the completion task is conditioned not just on the target context, but also on another source image from which the desired content is selected. Such a conditional image completion approach is also addressed by the work of Dolhansky and Ferrer~\shortcite{dolhansky2017eye}. However, their system focuses on the very specific task of copy-pasting eyes on face images only. Probably the most related previous work on this task is FaceShop~\cite{Portenier18}. Their system features a copy-paste mode that allows to cut facial features from a source image and paste it seamlessly into a target image. However, unlike their system, our approach is not restricted to face images, as we will show in this work. Moreover, our technique enables copy-pasting of features like textures and low-contrast content, which is not possible with their system.

The core component of our system is a deep convolutional neural network (CNN) that is trained end-to-end to perform copy-pasting. Our framework can be trained on any image dataset without the need of additional information such as labels or paired images. We show that our approach produces high-quality copy-paste results for a wide range of difficult examples where previous methods fail. We evaluate our technique on two completely different image datasets and demonstrate its effectiveness on a wide range of examples. In summary, we make the following contributions:
\begin{itemize}
 \item a novel technique to generate training data that is suitable to train a CNN on the task of copy-pasting objects from one image to another, given an arbitrary image dataset without additional information as input,
 \item an end-to-end trained system that synthesizes high-resolution, high-quality, and coherent copy-paste results,
 \item a framework that outperforms the previous state of the art technique on face images, and
 \item we demonstrate its effectiveness beyond faces on a diverse image dataset.
\end{itemize}

\section{Related Work}
In this section, we discuss previous work that is most related to our technique. A comprehensive survey of image editing techniques would exceed the scope of this work, therefore we will focus on approaches that are highly related either technically or conceptually. The following paragraphs are divided into two categories: deep image completion techniques and traditional copy-paste approaches. Image completion (sometimes named deep image inpainting) considers the problem of filling a missing region in an input image, either by using additional inputs or solely based on the region context. Recent advances in deep learning inspired many recent techniques to solve this task by leveraging deep neural networks. Traditional copy-paste approaches (also known as image harmonization techniques) try to solve the task of seamlessly blending a region in a source image into a target image. We focus on this task in our work, borrowing techniques from deep image completion approaches to produce photo-realistic copy-paste results.

\subsection{Traditional Copy-Paste Techniques}
Poisson image editing~\cite{Perez2003} is a technique that performs copy-paste in image gradients domain. This technique works very well in examples where the shading mismatch between source and target is moderate, but it causes distracting blending artifacts if the discrepancy is too severe. Moreover, it struggles with geometric inconsistencies, since it is not suitable to ignore mismatching features or to hallucinate missing content, as we will show in Section~\ref{sec:results}. Another traditional technique extends Poisson image editing by finding optimal seams that reduce artifacts using graph-cut optimization~\cite{Boykov2001}, before applying Poisson blending to blend the source and target image~\cite{Agarwala2004}. While their system mitigates blending artifacts by reducing the shading mismatch, it features the same problems as Poisson image editing in terms of geometric inconsistencies.

\subsection{Deep Image Completion}
Many recent techniques leverage CNNs to be trained on the task of image completion. Pathak et al.~\shortcite{pathakCVPR16context} successfully tackle the problem by training a Generative Adversarial Network (GAN)~\cite{goodfellow2014} in combination with a pixel-wise reconstruction loss. GANs are trained by leveraging an auxiliary discriminator network that acts as a loss function. The image completion network tries to fool the discriminator network, which itself tries to discriminate genuine from synthesized images. The technique has further been improved by using two discriminator networks for two different scales~\cite{Li_2017_CVPR, IizukaSIGGRAPH2017}, which leads to greater synthesis quality. One obvious drawback of these techniques is that the user has no control on the completion process, since it is entirely determined by the image context.

Recently, researchers have proposed to leverage additional input information to guide the image completion process. Dolhansky et al.~\shortcite{dolhansky2017eye} developed a system that is able to inpaint missing eye regions into face images by incorporating another input image that provides exemplar eyes. They demonstrate the effectiveness on the application of opening eyes in a portrait image that features closed eyes. While their results are promising, the application is very limited to eye regions only. Portenier et al.~\shortcite{Portenier18} proposed a system that enables to copy-paste facial parts such as nose, mouth, or eyes by leveraging a sketch domain as copy-paste space and training a CNN to translate the sketch to a photo-realistic image. Their system produces exciting results, however, they demonstrate it only on face images. Moreover, their system cannot copy-paste features that are lost by the transformation to the ad hoc sketch domain, e.g., textural features or low-contrast images. In contrast to these techniques, our system is applicable beyond face images and enables copy-pasting of faint features such as textures. Another closely related technique is the deep image harmonization network by Tsai et al.~\shortcite{Tsai_2017_CVPR}. They follow a similar approach to train a CNN to do copy-paste, however, similar to the work by Yang et al.~\shortcite{Yang18}, their technique requires semantic label information to train their network, unlike our approach that works with any image dataset without additional information. Moreover, their system is trained to handle only discrepancies between source and target due to shading, whereas our system can also handle geometric inconsistencies such as background clutter, missing information or pose mismatches. Other techniques relax the copy-paste problem by training the CNN to choose the best matching region from a source image to do the inpainting~\cite{Zhao18}. While this approach is more flexible, automatically selecting a region in a source image limits the user's ability to control the result, leading to undesired copy-paste results in practice.

\section{Smart, Deep Copy-Paste}
\begin{figure}[t]
\centering
\includegraphics[width=1\linewidth]{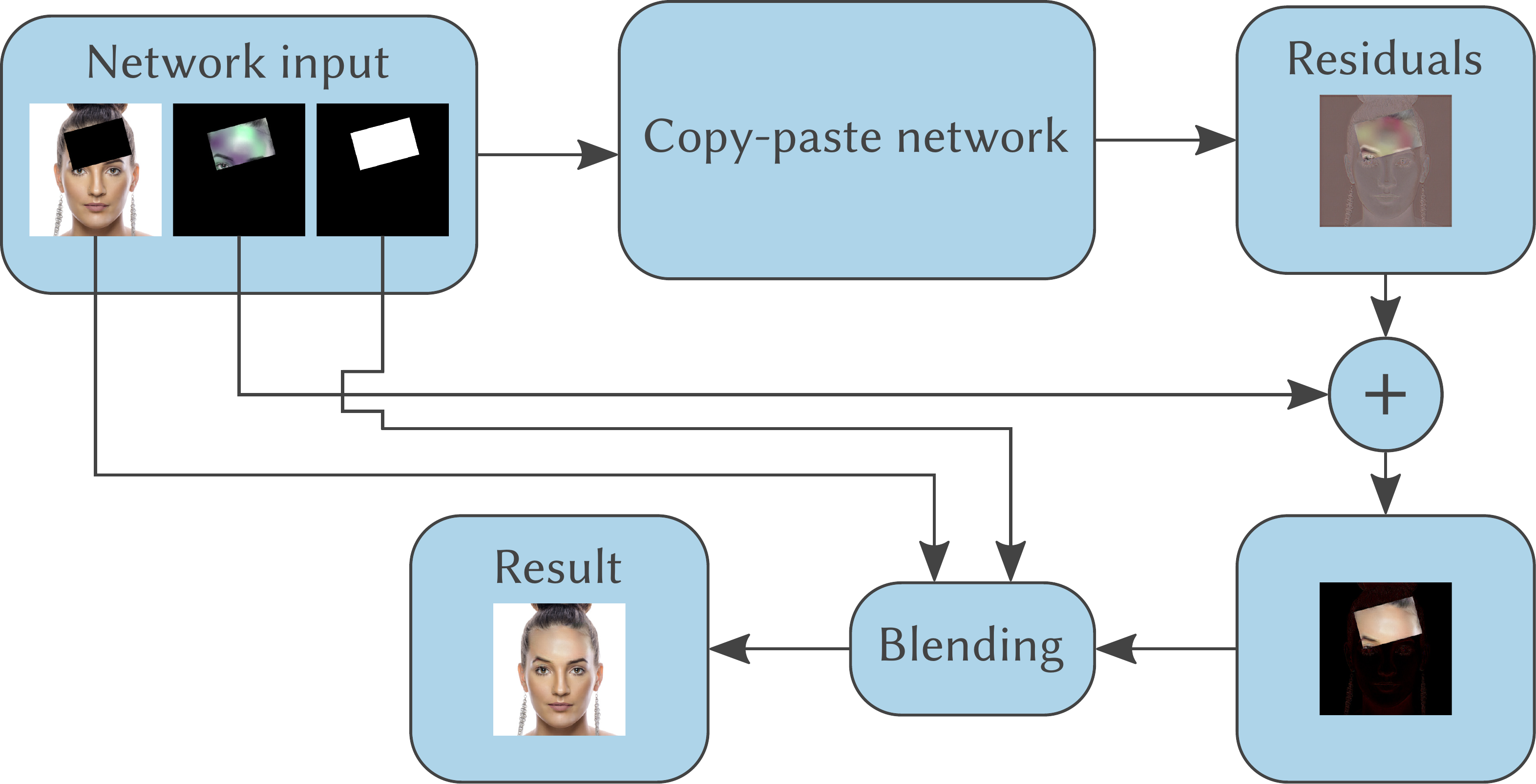}
\caption{System overview. The main component consists of a deep CNN that is trained on the task of copy-pasting. At runtime, the input to our network is a target image, a masked region from a source image, and a binary mask. The system then computes residuals that are added to the source image, before the source and target images are composited using the provided mask to form the final result.}
\label{fig:overview}
\end{figure}
In this section we introduce our smart copy-paste framework, which is implemented using a deep CNN. Figure~\ref{fig:overview} shows an overview of the proposed system. The main component consists of a CNN that is trained using a combination of conditional GAN loss and reconstruction loss. At runtime, the input to our network consists of a target image, a masked region from a source image and a binary mask. The system then computes residuals that are added to the source image, before the source and target images are composited using the provided mask to form the final result.

In Section~\ref{sec:training_data} we discuss our approach to produce suitable training data, given an arbitrary image dataset as input. This procedure is crucial to render realistic results without the need of additional labels. Next, we motivate and describe the network architecture that we use to implement our copy-paste system (Section~\ref{sec:net_arch}). Finally, we propose a training procedure in Section~\ref{sec:training} that allows the training of our model end-to-end using suitable training data.

\subsection{Training Data}
\label{sec:training_data}
The task of copying content from one image and seamlessly paste it into another image is difficult to train. Ideal training data consists of paired source and target images $(I_S, I_T)$, where one region in $I_T$ semantically fits into another region in $I_S$. In addition, a binary mask $M$ that indicates the respective regions in both images is required. Finally, a ground truth image is needed for each training pair in order to provide a meaningful loss signal during training. While the generation of suitable image pairs could be achieved by manually selecting and labeling appropriate images, the generation of ground truth images to compile a large-scale dataset suitable to train a deep CNN is hardly possible in practice. To overcome these issues, we propose a novel training data generation procedure that works with any image dataset without additional labeling.

The core idea of our approach is to automatically produce image pairs $(I_S, I_T)$ procedurally, using a single image as input. Given a training image $I_T$, we cut out a random region, called \textit{content}, transform it using a transformation $T$, and paste it back into $I_T$. During training, the network tries to recover the original image, i.e., it tries to undo the transformation $T$ on the content region. Besides the obvious advantage of requiring only a single image instead of a suitable image pair, this approach also enables training with completely random masks, and it provides a ground truth image for each training sample, namely the original image itself. However, finding a suitable $T$ such that $T(I_T) = I_S$ is crucial: if the transformation is too simple and close to the identity, the task for the network becomes trivial and it will overfit to the training data and fail on real-world copy-paste examples, where $I_S$ and $I_T$ typically consist of two different images. On the other hand, if $T$ is too strong, the task becomes too difficult and the network will degenerate to unconditional inpainting, effectively ignoring the input $I_S$.

We therefore need to design $T$ carefully, ideally simulating real-world copy-paste image pairs. When considering copy-paste examples that we would like to be solved by our system, we observe that several discrepancies may occur between $I_S$ and $I_T$. There might exist color mismatches between source and target, based on inconsistencies in shading or background colors. Moreover, the content may feature inconsistent geometry, for example the pose of the object to paste can be slightly wrong. In addition, roughly cutting an object from a source image may include unrelated background clutter that should be ignored during copy-paste. These observations inspire our design of the transformation $T$, which consists of both shading and geometric transformations, i.e., $I_S = T(I_T) = (T_\text{shading} \circ T_\text{geometric})(I_T)$. Figure~\ref{fig:augmentations_example} for an example of $T$.

\subsubsection{Shading Adjustments}
\label{sec:shading_adjustments}
\begin{figure}[t]
\centering
\includegraphics[width=1\linewidth]{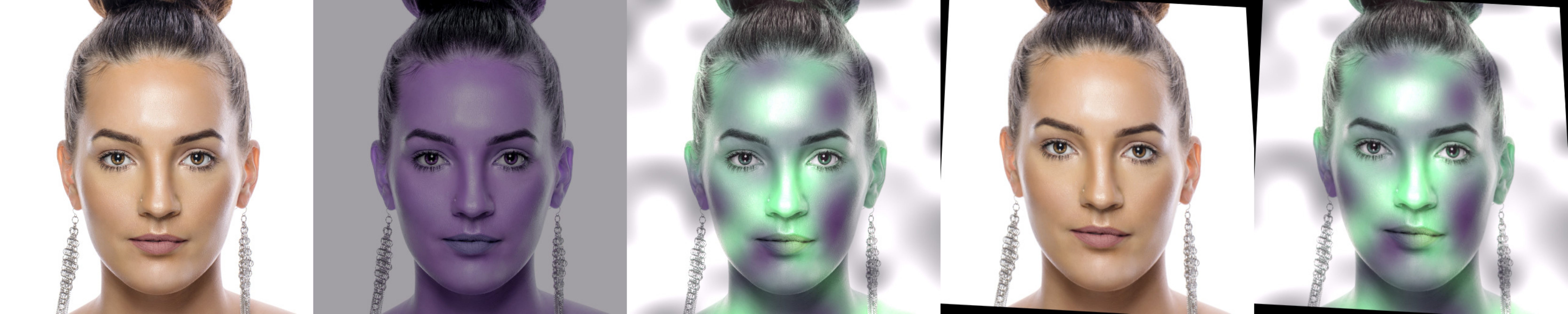}
\caption{Example of our proposed transformation $T$. From left to right: original image $I$, color transformation $T_{\text{color}}(I)$, shading transformation $T_{\text{shading}}(I)$, geometric transformation $T_{\text{geometric}}(I)$, and the complete final transformation $T$.}
\label{fig:augmentations_example}
\end{figure}

To mimic shading and other color mismatches, we first define a color transformation $T_\text{color}$ that randomly changes brightness, contrast, hue, and saturation of image $I$, i.e., 
\begin{equation}
\label{eq:color_transform}
T_\text{color}(I) = (T_s \circ T_h \circ T_c \circ T_b)(I),
\end{equation}
where $T_s(I)$ adjusts the saturation by converting $I$ from RGB to HSV, scaling the S-channel by factor $k_1\sim\mathcal{U}(0.5, 1.5)$, and converting the image back to RGB. Similarly, $T_h$ adjusts the hue by adding a bias $\Delta_1\sim\mathcal{U}(-0.5, 0.5)$ to the H-channel. $T_c$ adjusts the contrast for each channel independently by scaling the zero-centered pixel values: 
\begin{equation}
\label{eq:contrast_transform}
T_c(I) = ((I_1 - m_1) \lambda_1 + m_1, (I_2 - m_2) \lambda_2 + m_2, (I_3 - m_3) \lambda_3 + m_3),
\end{equation}
where $m_i$ is the mean pixel value of color channel $i$ and $\lambda_i\sim\mathcal{U}(0.5, 1.5)$. Finally, $T_b$ adjusts the brightness by adding a single bias $\Delta_2\sim\mathcal{U}(-0.5, 0.5)$ to all color channels. See Figure~\ref{fig:augmentations_example} for an example of $T_{\text{color}}$.

Global transformations such as $T_\text{color}$ can easily be learned to be undone by a CNN by applying appropriate scaling and bias to the signal. In practice, more complicated mismatches between source and target image occur. The content patch may fit locally with the target image in one region, but another region may feature inconsistencies such that global adjustments do not resolve the issue. To simulate such locally varying color mismatches, we propose a more sophisticated procedure to simulate locally varying shading adjustments. This step is crucial, as we will show in Section~\ref{sec:results}. Given input image $I$, we first compute two independent color transformations $I_1 = T_\text{color}(I)$ and $I_2 = T_\text{color}(I)$. Next, we compute a random mixing mask $M_\text{mix}$ that is used to fuse the two images $I_1$ and $I_2$ to form the final image that features locally varying shading adjustments:
\begin{equation}
\label{eq:shading_transform}
T_\text{shading}(I) = \max(\min(I_1 M_\text{mix} + I_2 (1 - M_\text{mix}), 1), 0).
\end{equation}
To create $M_\text{mix}$, we first sample a single-channel image of resolution $10 \times 10$ with salt-and-pepper noise. Next we use bilinear upsampling to scale the noise image to the resolution of $I$. See Figure~\ref{fig:augmentations_example} for an example of $T_\text{shading}$.

\subsubsection{Geometric Adjustments}
\label{sec:geometric_adjustments}
So far we considered how to handle color mismatches. In real-world copy-paste examples, geometric inconsistencies between source and target images occur. To address such geometric mismatches we propose to apply random homography transformations to the training image. For images of planar objects, homographies encode all possible pose transformations. This is no more true for more general, non-planar objects, but we find that if the homography is sufficiently close to the identity, the error is negligible. In addition to pose mismatches, applying homographies enables our system to become robust to other mismatches. Since we apply $T$ before we cut and paste a region during training, we introduce two other types of inconsistencies that also occur in practice. First, some content might get lost by applying the homography, and the system is encouraged to learn to invent missing content, based on the pasted content and the context. In addition, image features might appear twice after applying the homogaphy, once in the context region and again in the pasted content. The network needs to learn to ignore such content in order to complete the task. This makes our system robust to both missing content and clutter, as we will show in Section~\ref{sec:results}.

To apply a random homography, we first add random offsets $o_i\sim\mathcal{U}(-\sigma, \sigma)$ to the four image corner pixel coordinates $p_i$. Next, we compute the respective homography $H$ by solving the linear system consisting of the point correspondences $(p_i, p_i + \sigma_i)$ using least squares. Therefore, our proposed geometric transformation is $T_\text{geometric}(I) = H(I)$. See Figure~\ref{fig:augmentations_example} for an example of $T_\text{geometric}$.

Similar to \cite{Portenier18}, we use randomly rotated rectangular masks $M$ with random size and position. Each training sample consists of a ground truth image $I$, a target image $I_T = I (1 - M)$, a source image $I_S = T(I) M$, and a copy-paste mask $M$.

\subsubsection{Datasets}
To evaluate the effectiveness of our system, we train it on two different datasets. First, we use the high-resolution face dataset from \cite{Portenier18}, which allows a qualitative comparison to FaceShop. This dataset consists of 21k face images of $512 \times 512$ resolution. Next, we train our system on the Cityscapes dataset
\cite{Cordts2016Cityscapes} to demonstrate its effectiveness on non-face images. This dataset consists of 5k street view images captured by a camera mounted on top of a car driving through different cities. The images feature a resolution of $1024 \times 2048$ pixels, and we downsample the images for training to a resolution of $512 \times 1024$ pixels.

\subsection{Network Architecture}
\label{sec:net_arch}
In this section, we motivate and describe the network architecture that enables our system to render high-quality copy-paste results. In FaceShop~\cite{Portenier18}, the authors propose a copy-paste system that leverages a sketch domain as copy-paste space to produce realistic renderings on face images. The main idea is to map a source image to the sketch domain, copy-paste a source region in this domain to a target image, and let a conditional completion network produce the final result. This ad hoc sketch domain comes with an obvious drawback: features like textures or low contrast content are lost by mapping an image to the sketch domain. Hence, the system cannot be applied to copy-paste such image features. Moreover, the proposed sketch domain may not work reasonably well on arbitrary image datasets, and the authors only demonstrate its effectiveness on face images. Our core idea is therefore to learn a copy-paste space that is more suited for the task than the ad hoc sketch domain. A promising idea is to train a separate encoder network that maps the source image to a dedicated copy-paste space, and the source content is then cut out in this copy-paste domain and fed into another copy-paste network that computes the final result. The proposed encoder network can be trained simultaneously to the copy-paste network to learn an optimal representation for the copy-paste task. We experimented with such an explicit copy-paste space, but our experiments showed that we achieve better results by using a single copy-paste network without a dedicated copy-paste space encoder. Our single copy-paste network implicitly learns an internal representation that is better suited to do the copy-paste task, by leveraging both the source content as well as the target context to find a suitable representation that is invariant to the transformations introduced in Section~\ref{sec:training_data}, and in the same time keeps as much information as possible from the source image. Based on how well the copied content fits with the target context, our network maps the input to a more or less abstract representation.
Inspired by \cite{Portenier18}, we use an encoder-decoder architecture for our copy-paste network. In addition, we leverage an auxiliary discriminator network that acts as a loss function that is learned simultaneously to the copy-paste network. The input to both networks are images of resolution $512 \times 512$.

\paragraph{Copy-Paste Network}
The encoder part of the copy-paste network gradually downsamples the input tensor to a spatial resolution of $64 \times 64$ pixels, increasing the number of feature maps to 512. Each downsampling operation is implemented using a strided convolution layer with stride~2, followed by a non-strided convolution layer. Each convolutional layer is followed by a leaky ReLU nonlinearity with $\alpha = 0.2$, and a local response normalization layer \cite{karras2018progressive}, defined as:
\begin{equation}
\label{eq:lrn}
LRN(a_{x,y}) = \frac{a_{x,y}}{\sqrt{\frac{1}{N}\sum_{i=0}^{N-1}(a_{x,y}^i)^2 + \epsilon}},
\end{equation}
where $a_{x,y}^i$ is the activation of feature map $i$ at coordinate $(x,y)$, $N$ is the number of feature maps, and $\epsilon = 10^{-8}$ prevents division by zero.

The intermediate layers of the copy-paste network are implemented using seven dilated convolution layers with stride~1 and increasing dilation rate up to 16 in order to increase the receptive field. Each layer is again followed by a leaky ReLU nonlinearity and a LRN layer.

The decoder part of the copy-paste network is a mirrored version of the encoder part, gradually upsampling the activations back to the input resolution and decreasing the number of feature maps to three RGB channels. Each upsampling step is implemented using a transposed convolution layer, followed by a non-strided convolution layer. In addition, we add a noise addition layer after each transposed convolution layer, before evaluating the nonlinearity. This injection of stochastic information helps the network to fill in missing information, as we will show in Section~\ref{sec:results}. Inspired by~\cite{karras2018style}, we sample a single-channel image with per-pixel Gaussian noise for each layer, and broadcast it to the number of feature maps in the corresponding layer. The noise image is then added to each feature map, scaled by a learned per-channel variable. We find that this method produces better results than feeding a single noise image as input to the network.

In an U-Net~\cite{RFB15a} manner, we additionally use skip connections between all corresponding downsampling and upsampling layers by concatenating the respective feature channels, similar to~\cite{Portenier18}. All convolution layers use kernel size $3 \times 3$ and the transposed convolution layers feature kernel size $4 \times 4$.

\paragraph{Discriminator Network}
We borrow the discriminator architecture from \cite{Portenier18}. It consists of a global branch that consumes the entire copy-paste result and a local branch that focuses on the pasted region. Both branches are merged using a single linear dense layer that outputs a scalar value. For a detailed network architecture we refer to \cite{Portenier18}.

\subsection{Training Procedure}
\label{sec:training}
Next, we explain the training inputs and outputs and the loss function in detail. Finally, we propose important training details and the choice of hyperparameters that allow stable training.

\begin{figure}[t]
\centering
\includegraphics[width=1\linewidth]{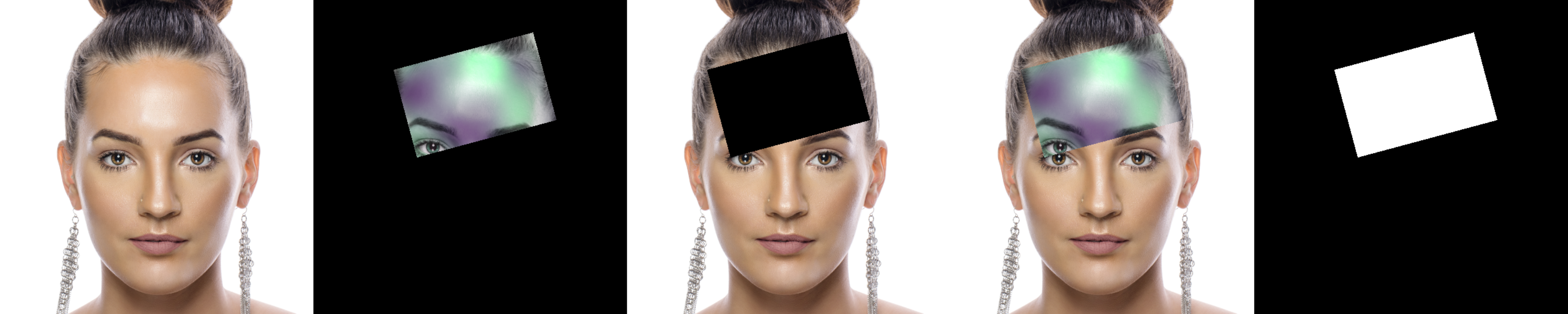}
\caption{Example of the proposed training input to the copy-paste network. From left to right: $I$, $T(I)M$, $I(1-M)$, $I(1-M) + T(I)M$, and $M$.}
\label{fig:training_sample}
\end{figure}
The training input $X$ to our copy-paste network $G$ is a $512 \times 512 \times 4$ tensor, composed of the target context, the transformed source content, and the copy-paste mask, i.e., $X = (I(1-M) + T(I)M, M)$, where $I$ is a random $512 \times 512$ crop of a training image and $M$ is a copy-paste mask of random position, size and rotation, as described in Section~\ref{sec:training_data}. See Figure~\ref{fig:training_sample} for an example. Since a lot of information is already present in the input $X$ and $G$ only has to make the pasted content fit to the context, we let $G$ learn only the residuals instead of the final image. We find that this approach produces better results and the network does not ignore the input content even with moderately strong transformations $T$, as opposed to learning the final image directly. In addition, we replace the image context with the genuine input context before feeding the final image $Y$ to the loss function, i.e., $Y = (G(X) + T(I))M + I(1-M)$. This encourages $G$ to focus on the content only, since we do not want the system to change the context. The input to the discriminator network $D$ consists of an RGB image, either a genuine image $I$ or a generated image $Y$. We use a conditional GAN, i.e., we concatenate the copy-paste input as additional information to the discriminator input, thus the overall input to $D$ is $(Y, I(1-M) + T(I)M, M)$.

As training loss function, we use a combination of WGAN-GP~\cite{gulrajani2017} and a pixel-wise reconstruction loss, with an additional regularization term to minimize the norm of the logits~\cite{karras2018progressive}. The WGAN-GP loss is defined as
\begin{equation}
\label{eq:wgangp}
L_{WGAN-GP} = \mathbb{E}[D(Y)] - \mathbb{E}[D(I)] + \lambda \mathbb{E} [(||\nabla_{I_u}D(I_u)||_2 - 1)^2],
\end{equation}
where $I_u$ is an uniformly sampled image along the straight line between $I$ and $Y$, and we set $\lambda = 10$. As reconstruction loss we use the pixel-wise $L_1$ distance, i.e.,
\begin{equation}
\label{eq:l1}
L_{rec}(Y) = \frac{1}{N}\sum_{i=0}^{N-1}|Y_i - I_i|,
\end{equation}
where $N$ is the number of image pixels. Our final training objective is therefore defined as
\begin{equation}
\label{eq:loss}
L = L_{rec} + \alpha L_{WGAN-GP} + \gamma \mathbb{E}[D(I)^2].
\end{equation}

In all our experiments, we set $\alpha = 10^{-4}$ and $\gamma = 10^{-3}$. We use ADAM optimizer~\cite{kingmaB14} with $\beta_1 = 0.0$ and $\beta_2 = 0.9$, and we use a constant learning rate of $2 \times 10^{-4}$ for both $G$ and $D$. For both datasets, we use batch size of 5 and train the networks for 250k iterations, which takes approximately two weeks on a Titan XP GPU. For the face image dataset, we set $\sigma = 15$. Since the Cityscapes dataset is harder to train, we found that setting $\sigma = 10$ leads to better results on this dataset.

\section{Results}
\label{sec:results}
In this section, we first show ablation studies that demonstrate the effectiveness of different design decisions in our framework (Section~\ref{sec:ablations}). Next, we show qualitative copy-paste results on the face images dataset with a comparison to the state of the art technique (Section~\ref{sec:face_results}). In Section~\ref{sec:cityscapes_results} we finally show results on a dataset that is more complex than face images.

\subsection{Ablation Studies}
\label{sec:ablations}
\begin{figure}[t]
\centering
\includegraphics[width=1\linewidth]{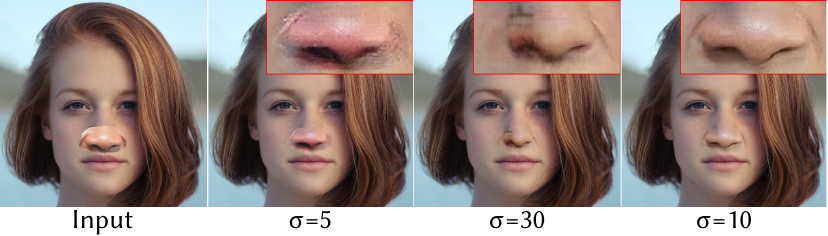}
\caption{Effect of different values for the offset parameter $\sigma$ in $T_{\text{geometric}}$. When choosing a value that is too small (second column), the network fails on real-world copy-paste examples and introduces visual artifacts (see zoomed crop). These artifacts do not occur on training data, which is a sign of overfitting. If $\sigma$ is too large (third column), the task becomes too difficult and the network partially ignores the source content, rendering a differently shaped nose in this example. Only the right amount of geometric transformation produces decent results that both resemble the source content and do not feature overfitting artifacts (last column).}
\label{fig:offset_values}
\end{figure}
We first show the effect of different choices for the parameter $\sigma$ that controls the strength of the geometric transformation $T_{\text{geometric}}$ (Section~\ref{sec:geometric_adjustments}). Choosing the right amount of geometric transformation is crucial for our system to produce high-quality copy-paste results, as demonstrated in Figure~\ref{fig:offset_values}. If the offset parameter is too small, the network overfits to the training data, noticeable as visual artifacts when feeding source content that comes from another image than the target context. When setting the offset too high, the copy-paste network tends to ignore the source content for difficult examples and degenerates to some sort of unconditional inpainting. Only when setting a reasonable value for the offset parameter, the system synthesizes high-quality results that resemble the input content without introducing visual artifacts. The same argument holds for the strength of the shading transformation $T_{\text{shading}}$. Inappropriate choices for the color transformation parameters either cause the network to become totally invariant to colors, making it impossible to copy-paste color features, or render the network unable to resolve examples with strong shading mismatches between source and target.

\begin{figure}[t]
\centering
\includegraphics[width=1\linewidth]{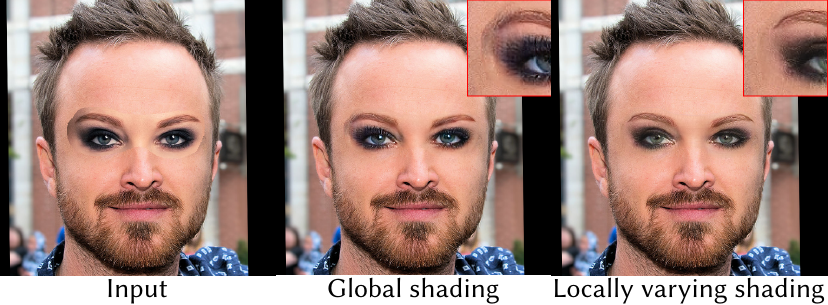}
\caption{Effect of locally varying shading adjustments. Using simple, global shading adjustments (second column) causes the network to fail on examples with locally varying shading mismatches. Our proposed locally varying shading adjustments lead to a copy-paste network that effectively solves such examples.}
\label{fig:global_vs_local_shading}
\end{figure}
In Figure~\ref{fig:global_vs_local_shading} we demonstrate the effect of the local shading adjustments proposed in Section~\ref{sec:shading_adjustments}. Using simple global shading adjustments, the copy-paste network fails to seamlessly blend source and target image in examples where the shading mismatch between source and target changes locally. When training the network with our proposed locally varying shading adjustments, the network is able to resolve such mismatches.

\begin{figure}[t]
\centering
\includegraphics[width=1\linewidth]{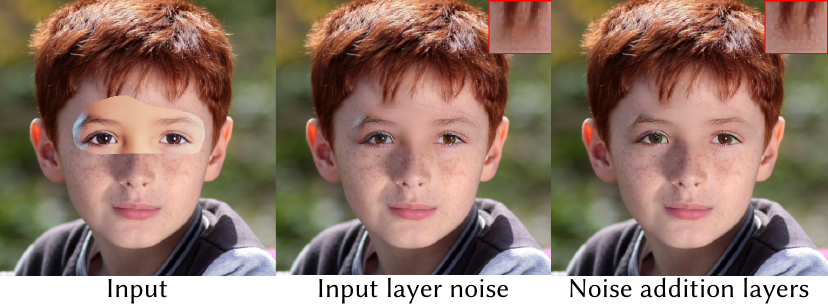}
\caption{Effect of noise addition layers. Feeding a single noise image to the input layer (second column) leads to less realistic synthesis of missing features, such as the strand of hair in this example. The use of noise addition layers (last column) produces more realistic results in these cases.}
\label{fig:noise_methods}
\end{figure}
Finally, we highlight the effectiveness of the proposed noise addition layers in the decoder part of the copy-paste network. In Figure~\ref{fig:noise_methods}, we compare the use of noise addition layers to feeding a single noise image to the input layer, as proposed by~\cite{Portenier18}. Noise addition layers enable the network to produce more realistic features in cases where content needs to be ``invented'' by the network based on the context. In contrast, feeding a single noise channel to the input layer that needs to be propagated through the entire network makes it more difficult to render realistic features.

\subsection{Face Images Dataset}
\label{sec:face_results}
\begin{figure*}[t]
\centering
\includegraphics[width=1\textwidth]{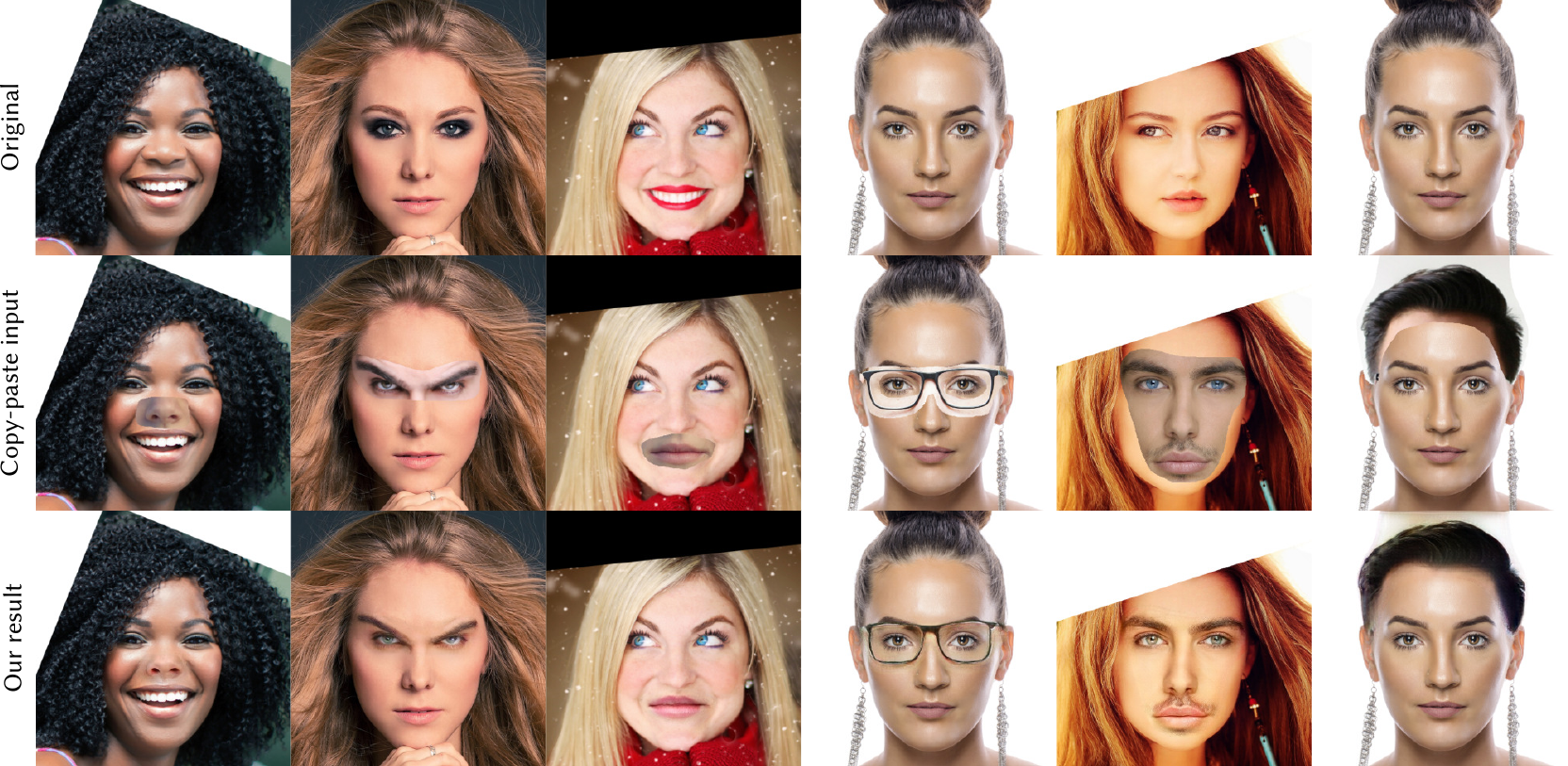}
\caption{Results on the face images dataset. Our system enables to copy-paste various facial features like eyes, nose, and mouth, but also face accessories such as glasses. Moreover, the framework can also be used to replace the entire face and hairstyle.}
\label{fig:face_results}
\end{figure*}
In Figure~\ref{fig:face_results} we show various results on face images, produced by our copy-paste network. All examples feature a resolution of $512 \times 512$ and show the final output of our framework, without any post-processing. The examples include copy-pasting of facial features like nose, mouth, eyes, or glasses. Moreover, we show examples of replacing the entire face or hairstyle, using source content patches that are significantly larger than the patches used for training.

\begin{figure*}[t]
\centering
\includegraphics[width=1\textwidth]{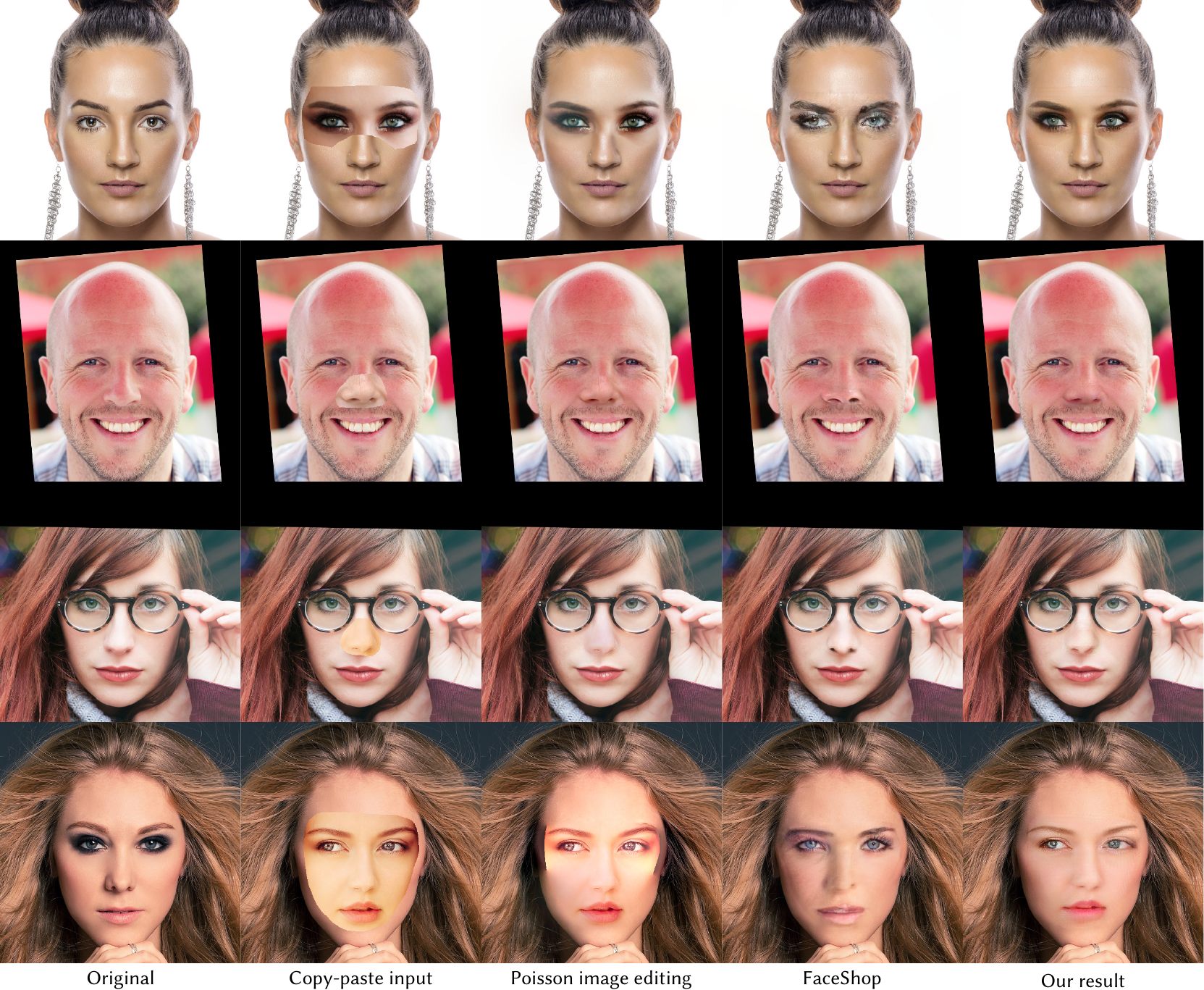}
\caption{Qualitative comparison of our copy-paste framework to FaceShop~\cite{Portenier18} and Poisson Image Editing~\cite{Perez2003}. Each row shows a different example. Our framework produces high-quality results, whereas Poisson image editing suffers from both shading and geometric inconsistencies. FaceShop clearly features a higher level of abstraction, the outcome is often less similar to the actual input features compared to our approach.}
\label{fig:face_comparisons}
\end{figure*}
Next, we compare our copy-paste network to the copy-paste mode in FaceShop~\cite{Portenier18}, a state of the art technique for copy-pasting facial features. In addition, we do a comparison to Poisson image editing~\cite{Perez2003}. Figure~\ref{fig:face_comparisons} shows our approach compared to the two baselines. In the first example, Poisson image editing does a good job at the right eye, but introduces subtle blending artifacts on the left eye due to the locally varying shading mismatch. FaceShop seems to be quite confused by this example. The strong eye makeup produces clutter edges in the sketch domain, and the network fails to interpret the edges in this example. Our approach manages to transfer both the eye and eyebrow geometry as well as the texture features. The second example shows rather successful results for all approaches, only Poisson image editing has difficulties to resolve the geometric mismatch. Interestingly, FaceShop produces a rather different nose shape than the input, probably due to the highly abstract sketch domain. In the third example, FaceShop actually falls back to unconditional image completion, since the edges of the input nose are so faint that the sketch domain ends up void. The last example shows a complete failure case for Poisson image editing. Interestingly, FaceShop also produces a result that is quite different from the input and somewhat blurry. Maybe this is a sign of overfitting, since FaceShop was not trained on such large masks.

\subsection{Cityscapes Dataset}
\label{sec:cityscapes_results}
\begin{figure*}[t]
\centering
\includegraphics[width=1\textwidth]{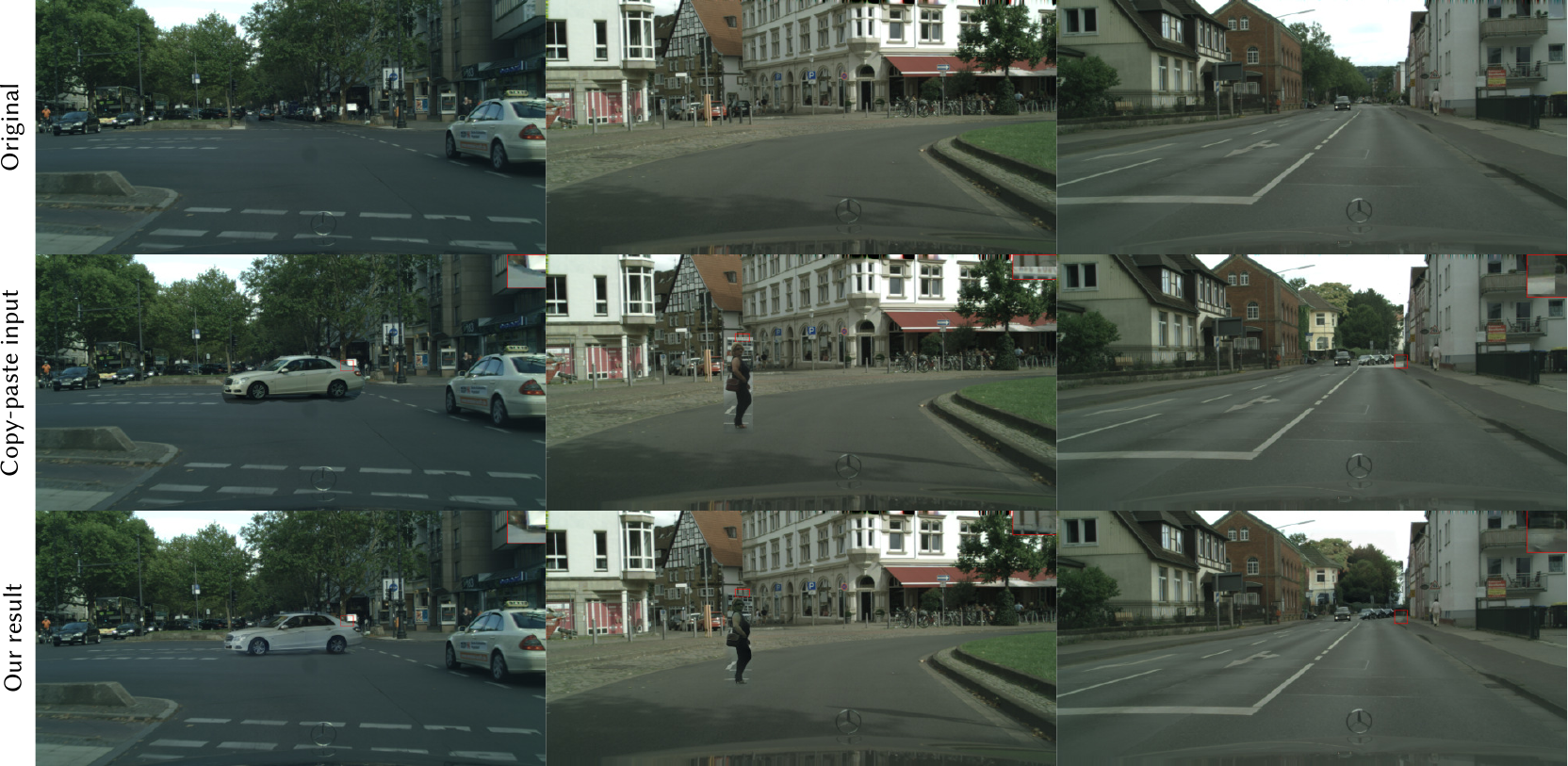}
\caption{Results on the Cityscapes images dataset. For each example (column), we show a zoomed crop at the top right corner that shows a region that is particularly difficult due to background clutter or shading mismatches.}
\label{fig:cityscapes_results}
\end{figure*}
Figure~\ref{fig:cityscapes_results} shows various results on the Cityscapes dataset. Since our copy-paste network is fully convolutional, we can apply it on arbitrary sized images after training. Note that all results are of resolution $1024 \times 2048$, even though the network was trained solely on $512 \times 512$ crops. Pasted regions feature an extent of up to 700 pixels (first column), significantly larger than the crops that our network consumed during training, demonstrating the generalization capability to higher resolutions, given appropriate training data. The results demonstrate that our approach also works on a more complex dataset, where the number of different objects and features are much more diverse than in the case of face images. It must be emphasized that the overall synthesis quality is worse compared to the face images dataset. We attribute this to the fact that copy-pasting on this dataset is more difficult than on face images, mainly due to background clutter when providing inaccurate masks. Moreover, the dataset features both heavy motion blur as well as tonemapping artifacts, limiting the ability of the network to synthesize higher-quality results.

\section{Conclusions}
In this work we propose a novel smart copy-paste framework that enables the synthesis of high-quality copy-paste results. The key ingredient of our system is a deep convolutional neural network trained end-to-end on the task of copy-pasting. Our key contribution is a novel, carefully designed training data generation procedure that works on any image dataset without additional label information. We demonstrate the effectiveness of our system on two high-resolution datasets, outperforming the state of the art in face image copy-pasting on many examples. Moreover, we show the application beyond faces and process images featuring up to two megapixels resolution, copying content as big as 700 pixels into a target image. In future work, we will leverage higher quality image datasets to produce even better results with our approach.

\bibliographystyle{unsrt}  
\bibliography{bibliography}  

\end{document}